\documentclass[twocolumn,showpacs,preprintnumbers,amsmath,amssymb,eqsecnum]{revtex4}
\usepackage{graphicx}% Include figure files
\usepackage{dcolumn}% Align table columns on decimal point
\usepackage{bm}% bold math

\begin{document}

%\preprint{APS/123-QED}

\title{Nodal Structure of Superconductors with Time-Reversal Invariance\\
and ${\bm Z}_2$ Topological Number}% Force line breaks with \\

\author{Masatoshi Sato}
\email{msato@issp.u-tokyo.ac.jp}
\affiliation{%
The Institute for Solid State Physics, The University of Tokyo, \\
Kashiwanoha 5-1-5, Kahiwa, Chiba, 277-8581, Japan
}%

\date{\today}% It is always \today, today,
             %  but any date may be explicitly specified

\begin{abstract}
A topological argument is presented for nodal structures of superconducting
 states with time-reversal invariance.
A generic Hamiltonian which describes a quasiparticle in superconducting
 states with time-reversal invariance is derived, and it is shown that
 only line nodes are topologically stable in single-band descriptions
 of superconductivity.
Using the time-reversal symmetry, we introduce a real structure and
 define topological numbers of line nodes.
Stability of line nodes is ensured by conservation of the
 topological numbers.
Line nodes in high-$T_{\rm c}$ materials, the polar state in $p$-wave paring
 and mixed singlet-triplet superconducting states are examined in detail.
\end{abstract}

% PACS, the Physics and Astronomy
                             % Classification Scheme.
%\keywords{Suggested keywords}%Use showkeys class option if keyword
                              %display desired
\maketitle

\section{Introduction}
Nontrivial nodal structures are one of the most noticeable
features of unconventional superconducting gap functions.
Conventionally, the nodal structures have been investigated by power law
behaviours of the temperature dependences of the specific heat, the NMR
relaxation rates and so on \cite{Leggett, SU}. 
Although the power law behaviours give a hint of the nodal structures,
they can not give the details such as the number and the
position of nodes.  
Recently, it has been increasingly clear that the angular-controlled
measurements of thermal conductivity and specific heat in a vortex state
are useful probes to determine the details of the nodal structures
\cite{Volovik, VHCN, WM}.
Various unexpected nodal structures of superconductors have
been found in this method \cite{ABRGT, ITYMSSFYSO, IYMSSO, IYSM,
IKNMWNTTM, INGMOSSTM, IKMBYSO, TNMMI, TSNMMI, DMYM, DMM, ASSSOMM, PSCKL,
PSCKL2, PCSBVTCKLC}.
This makes urgent the need for a better theoretical understanding of
nodal structures of superconductors.

The purpose of this paper is to reveal theoretically a generic nodal
structure of superconducting states with time-reversal invariance.
We use two different methods to investigate the nodal structures.
The first one is based on a universal Hamiltonian we will construct, and the
other is an argument using topological numbers in momentum space.
Unless stated explicitly, our results do not rely on any
specific symmetry of superconducting states except the time-reversal
symmetry. 

Topological arguments in momentum space are powerful tools to
investigate a nonperturbative aspect of quantum theories
\cite{NN,TKNN,Berry,Kohmoto}. 
For superconducting states with broken time-reversal invariance, 
a topological characterization of the nodal structures was given in
Refs.\cite{Volovik2,Volovik3,HRK}. 
On the other hand, we will present here a topological characterization
of nodal structures of superconducting states with time-reversal
invariance. 
We will introduce novel topological numbers and discuss stability of
the nodal structures.

The organization of this paper is as follows. In
Sec.\ref{sec:hamiltonian}, we derive a generic Hamiltonian describing a
quasiparticle in superconducting states with time-reversal invariance.
The quasiparticle spectra are given and their nodal structures are discussed.
In Sec.\ref{sec:example}, we examine stability of line nodes for
high-$T_{\rm c}$ materials, the polar state in $p$-wave paring, and mixed
singlet-triplet states, respectively.
In Sec.\ref{sec:topological_number}, a real structure is introduced and
novel ${\bm Z}_2$ topological numbers of line nodes are defined. 
We calculate the ${\bm Z}_2$ topological numbers for
high-$T_{\rm c}$ materials, the polar state in $p$-wave paring, and mixed
singlet-triplet states.
It is shown that the existence of the topological numbers ensures
stability of the line nodes.
Conclusions and discussions are given in Sec.\ref{sec:conclusion}.

\section{Hamiltonian with time-reversal symmetry and nodal structure}
\label{sec:hamiltonian}

We derive here a generic Hamiltonian which describes a quasiparticle
in superconducting states with time-reversal invariance.
Let us consider a single-band description of superconducting state
and start with the following Bogoliubov-de Gennes type Hamiltonian,
\begin{eqnarray}
H=\frac{1}{2}\sum_{\bm k}{\bm c}({\bm k})^{\dagger}H({\bm k}){\bm c}({\bm k}),
\end{eqnarray}
where 
${\bm c}({\bm k})$
is a four component notation of electron creation and annihilation
operators in momentum space,
\begin{eqnarray}
{\bm c}({\bm k})=
\left(
\begin{array}{c}
c_{\sigma}({\bm k})\\
c_{\sigma}(-{\bm k})^{\dagger}
\end{array}
\right),
\quad
(\sigma=\pm),
\end{eqnarray}
and $H({\bm k})$ is a $4\times 4$ Hermitian matrix which will be
determined below. 
Since ${\bm c}({\bm k})$ satisfies 
\begin{eqnarray}
{\bm c}(-{\bm k})^{*}=\Gamma {\bm c}({\bm k}), 
\quad
\Gamma=
\left(
\begin{array}{cc}
0 &1 \\
1 &0
\end{array}
\right),
\end{eqnarray}
we can set the matrix $H({\bm k})$ to satisfy the following
equation,
\begin{eqnarray}
\Gamma H({\bm k})\Gamma =-H(-{\bm k})^{*}.
\label{eq:particle-hole}
\end{eqnarray}
We also demand that the Hamiltonian is time-reversal invariant.
The time-reversal operation ${\cal T}$ is defined as
\begin{eqnarray}
{\cal T}{\bm c}({\bm k})=\Theta {\bm c}(-{\bm k})^{*}, 
\quad
\Theta=
\left(
\begin{array}{cc}
i\sigma_2 &0 \\
0& i\sigma_2
\end{array}
\right),
\end{eqnarray}
where $\sigma_i$'s $(i=1,2,3)$ are the Pauli matrices.
Time-reversal invariance of $H$ implies
\begin{eqnarray}
\Theta H({\bm k})\Theta^{-1}=H(-{\bm k})^{*}. 
\label{eq:t-invariance}
\end{eqnarray}
We solve Eqs.(\ref{eq:particle-hole}) and (\ref{eq:t-invariance}) to
obtain a generic Hamiltonian of a quasiparticle in
superconducting states with time-reversal invariance.

To solve Eq.(\ref{eq:t-invariance}), we rewrite the 16 components of
$H({\bm k})$ in terms of the identity matrix, 5 Dirac matrices
$\Gamma_a$ and 10 commutators $\Gamma^{ab}=[\Gamma_a,\Gamma_b]/(2i)$
\cite{KM}, 
\begin{eqnarray}
H({\bm k})=h_0({\bm k})1+\sum_{a=1}^{5}h_a({\bm k}){\Gamma_a}
+\sum_{a<b=1}^{5}h_{ab}({\bm k})\Gamma_{ab}. 
\label{eq:hamiltonian}
\end{eqnarray}
Here $h_0({\bm k})$, $h_{a}({\bm k})$'s and $h_{ab}({\bm k})$'s are real
functions of ${\bm k}$.
We choose the following representation of $\Gamma_a$,
\begin{eqnarray}
&&\Gamma_1=
\left(
\begin{array}{cc}
0 &1 \\
1 &0
\end{array}
\right), 
\quad
\Gamma_2=
\left(
\begin{array}{cc}
1 &0 \\
0 &-1
\end{array}
\right), 
\nonumber\\
&&\Gamma_3=
\left(
\begin{array}{cc}
0&-i\sigma_1 \\
i\sigma_1 &0
\end{array}
\right), 
\quad
\Gamma_4=
\left(
\begin{array}{cc}
0 &-i\sigma_2 \\
i\sigma_2 &0
\end{array}
\right), 
\nonumber\\
&&\Gamma_5=
\left(
\begin{array}{cc}
0 &-i\sigma_3 \\
i\sigma_3 &0
\end{array}
\right).
\end{eqnarray}
In this representation, we obtain 
\begin{eqnarray}
\Theta \Gamma_a \Theta^{-1}=\Gamma_a^{*},
\quad
\Theta \Gamma_{ab}\Theta^{-1}=-\Gamma_{ab}^{*}. 
\end{eqnarray}
Therefore, Eq.(\ref{eq:t-invariance}) is satisfied if we have
\begin{eqnarray}
h_0({\bm k})=h_0(-{\bm k}), \quad
h_{a}({\bm k})=h_{a}(-{\bm k}), 
\end{eqnarray}
and
\begin{eqnarray}
h_{ab}({\bm k})=-h_{ab}(-{\bm k}). 
\end{eqnarray}
In addition to Eq.(\ref{eq:t-invariance}), 
we impose Eq.(\ref{eq:particle-hole}) on
Eq.(\ref{eq:hamiltonian}). 
Using $\Gamma=\Gamma_1$ and the commutation relation of $\Gamma_{a}$'s,
we find that the following 8 functions should be identically zero,
\begin{eqnarray}
&&h_0({\bm k})=h_1({\bm k})=h_3({\bm k})=h_5({\bm k})=0,
\nonumber\\
&&h_{13}({\bm k})=h_{15}({\bm k})=h_{24}({\bm k})=h_{35}({\bm k})=0.
\end{eqnarray}
Therefore, the Hamiltonian contains 16-8=8 independent functions. 
It is convenient to introduce a new notation of the remaining functions,
\begin{eqnarray}
&&h_2({\bm k})=\epsilon({\bm k}),
\quad
h_4({\bm k})=-\psi({\bm k}),
\nonumber\\
&&(h_{45}({\bm k}),h_{14}({\bm k}),h_{34}({\bm k}))={\bm g}({\bm k}),  
\nonumber\\
&&(h_{25}({\bm k}),h_{12}({\bm k}),-h_{23}({\bm k}))={\bm d}({\bm k}).
\end{eqnarray}
All of them are real functions.
Here $\epsilon({\bm k})$ and $\psi({\bm k})$ are even functions of ${\bm k}$,
and ${\bm g}({\bm k})$ and ${\bm d}({\bm k})$ are odd functions of ${\bm
k}$.
In terms of them, $H({\bm k})$ is written as 
\begin{eqnarray}
H({\bm k})= 
\left(
\begin{array}{cc}
\epsilon({\bm k})+{\bm g}({\bm k})\cdot{\bm \sigma} & \Delta({\bm k})\\
\Delta({\bm k})^{\dagger} &-\epsilon({\bm k})
+{\bm g}({\bm k})\cdot{\bm \sigma}^{*}
\end{array}
\right), 
\label{eq:hamiltonian2}
\end{eqnarray}
where $\Delta(\bm k)$ is defined by
\begin{eqnarray}
\Delta({\bm k})=i\psi({\bm k})\sigma_2
+i{\bm d}({\bm k})\cdot{\bm \sigma}\sigma_2. 
\end{eqnarray}
Now physical meanings of these functions are evident.
The function $\epsilon({\bm k})$ is a band energy of electrons
measured relative to the chemical potential $\mu$, and $\Delta({\bm k})$
is a gap function of a superconducting state. 
($\psi({\bm k})$ and ${\bm d}({\bm k})$ represent
the spin-singlet and spin-triplet gaps, respectively).
The function ${\bm g}({\bm k})$ is a parity breaking term in the
normal state.
For example, the Rashba term is represented by this. 

The quasiparticle spectra $E({\bm k})$ in the superconducting state are
given by the eigenvalues of $H({\bm k})$.
The eigenvalues can be obtained straightforwardly, and the resultant
spectra $E({\bm k})$ are
\begin{eqnarray}
&&E({\bm k})=\nonumber \\
&&\hspace{-3ex}\pm \sqrt{\epsilon^2+\psi^2+{\bm g}^2+{\bm d}^2
\pm 2\sqrt{(\epsilon {\bm g}+\psi{\bm d})^2+({\bm g}\times{\bm d})^2} } 
\nonumber\\
&&\hspace{-3ex}\equiv \pm E_{\pm}({\bm k}).
\label{eq:quasiparticle_energy}
\end{eqnarray}
Zeros of $E_{\pm}({\bm k})$ determine the nodal structure of the
superconducting state.
First, it can be shown that $E_{+}({\bm k})$ has no zero in general;
For $E_{+}({\bm k})$ to be zero, we have (at least) 
\begin{eqnarray}
\epsilon({\bm k})^2+\psi({\bm k})^2+{\bm g}({\bm k})^2+{\bm d}({\bm k})^2=0. 
\end{eqnarray}
This leads to 8 conditions, 
\begin{eqnarray}
\epsilon({\bm k})=\psi({\bm k})={\bm g}({\bm k})={\bm d}({\bm k})=0,
\end{eqnarray}
which can not be met generally in three dimensional momentum space. 
Even if $E_{+}({\bm k})$ has an accidental zero at some ${\bm k}$, we
can easily remove this by a small deformation of the Hamiltonian.
Therefore, the zero of $E_{+}({\bm k})$ is topologically unstable.

Let us now consider the condition $E_{-}({\bm k})=0$.
The Hermitian property of $H({\bm k})$ ensures that the eigenvalue
$E_-({\bm k})$ is real.  
Therefore, we have the inequality,
\begin{eqnarray}
\epsilon^2+\psi^2+{\bm g}^2+{\bm d}^2
\ge 2\sqrt{(\epsilon{\bm g}+\psi{\bm d})^2+({\bm g}\times{\bm d})^2}.
\label{eq:ine}
\end{eqnarray}
$E_{-}({\bm k})$ is zero when the equality in
(\ref{eq:ine}) is attained. 
We rewritten this as
\begin{eqnarray}
&&\epsilon^2+\psi^2+{\bm g}^2+{\bm d}^2
\nonumber\\
&&\hspace{-7ex}\ge 2\sqrt{
\epsilon^2{\bm g}^2+\psi^2{\bm d}^2
+{\bm g}^2{\bm d}^2+\epsilon^2\psi^2
-({\bm g}\cdot {\bm d}-\epsilon\psi)^2}. 
\label{eq:ine2}
\end{eqnarray}
With fixed $\epsilon^2$, $\psi^2$, ${\bm g}^2$ and ${\bm d}^2$, we can
maximize the right hand side of (\ref{eq:ine2}) at ${\bm
g}\cdot{\bm d}=\epsilon\psi$. 
Then if we assume ${\bm g}\cdot{\bm d}=\epsilon\psi$, the equality in
(\ref{eq:ine2}) is attained if we have
$\epsilon^2+{\bm d}^2=\psi^2+{\bm g}^2$.  
Therefore, $E_{-}({\bm k})$ becomes zero when
\begin{eqnarray}
&&{\bm g}({\bm k})\cdot{\bm d}({\bm k})=\epsilon({\bm k})\psi({\bm k}),
\label{eq:node1}\\
&&
\epsilon({\bm k})^2+{\bm d}({\bm k})^2=\psi({\bm k})^2+{\bm g}({\bm k})^2. 
\label{eq:node2}
\end{eqnarray}
We can show that these two equations are the necessary and
sufficient condition for $E_{-}({\bm k})=0$;
If the equality in (\ref{eq:ine2}) is attained under
a condition different from Eqs.(\ref{eq:node1}) and (\ref{eq:node2}),
the right hand side of Eq.(\ref{eq:ine2}) exceeds the left hand side by
imposing Eq.(\ref{eq:node1}).  
This contradicts the inequality (\ref{eq:ine2}).
Therefore, only Eqs.(\ref{eq:node1}) and ({\ref{eq:node2}}) lead to
$E_{-}({\bm k})=0$.   

The two equations (\ref{eq:node1}) and (\ref{eq:node2}) define two
surfaces in three dimensional momentum space.
They intersect in a line in general, and the intersection line gives
zeros of $E_{-}({\bm k})$.
Small deformations of the Hamiltonian change the surfaces slightly, but
the intersection line does not vanish.
This means that {\it for superconductors described by $H({\bm k})$, we
have a line node in general and the line node is topologically stable.}
On the other hand, a point node is accidental and it can be removed by a
small deformation of the Hamiltonian.

It is interesting to compare this result with that of group theoretical
analyses. 
Superconducting states are known to be classified by using
a group theoretical technique and the generalized Ginzburg-Landau theory
\cite{VG, VG2, Blount, UR, UR2, SU}.
The classification shows that point nodes exist in a superconducting
state with time-reversal invariance. 
For example,
Table VI in Ref.\cite{SU} shows that the gap function 
in the
$\Gamma^{-}_2$ representation of the tetragonal group ${\rm D_{4h}}$ has
the form
\begin{eqnarray}
{\bm d}({\bm k})=\hat{{\bm x}}k_y-\hat{\bm y}k_x.
\end{eqnarray}
This gap function preserves the time-reversal symmetry, and it has two
point nodes on $k_z$ axis. 
This result does not contradict ours.
Since the existence of the point nodes are due to a symmetry
property of the $\Gamma_2^{-}$ representation, we can remove these
nodes by a small perturbation which are not the $\Gamma_2^{-}$
representation.
Indeed, they are removed by the deformation 
\begin{eqnarray}
{\bm d}({\bm k})\rightarrow {\bm d}({\bm k})+\eta\hat{\bm z}k_z, 
\quad
(\eta<<1) 
\end{eqnarray}
where $\eta\hat{\bm z}k_z$ is the $\Gamma_1^{-}$ representation of 
${\rm D_{4h}}$.
Generally, the group theoretical method does not ensure the stability
of the nodal structure against any perturbations which change the
representation of the group.
On the other hand, our results are robust against any time-reversal
invariant perturbations.

\section{Examples}
\label{sec:example}

Using Eqs. (\ref{eq:node1}) and (\ref{eq:node2}), we can examine stability
of line nodes.
Here we consider three examples, 1) high-$T_{\rm c}$ materials, 2) the polar
state in $p$-wave pairing, and 3) mixed singlet-triplet
superconducting states.

\subsection{high-$T_{\rm c}$ materials}
\label{sec:ex_high_tc}

\begin{figure}[h]
\begin{center}
\includegraphics[width=5cm]{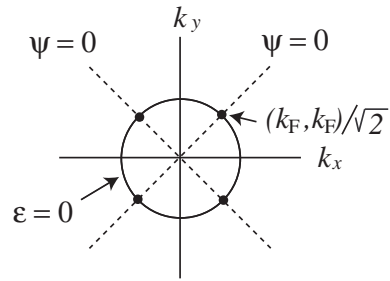}
\caption{Line nodes in high-$T_{\rm c}$ superconductors.}
\label{fig:high_tc}
\end{center}
\end{figure}

In high-$T_{\rm c}$ materials, the parity is conserved, thus we have
\begin{eqnarray}
{\bm g}({\bm k})={\bm d}({\bm k})=0. 
\end{eqnarray} 
The band energy $\epsilon({\bm k})$ mainly depends on $k_x$ and $k_y$,
and the Fermi surface given by $\epsilon({\bm k})=0 $ is two
dimensional. The gap function is
\begin{eqnarray}
\psi({\bm k})=\Delta_0(k_x^2-k_y^2).
\end{eqnarray}
In this state, $E_{+}({\bm k})=E_{-}({\bm k})$, and 
we have four line nodes on the
Fermi surface at ${\bm k}_0=(\pm k_{\rm F},\pm k_{\rm
F})/\sqrt{2}$. (See Fig.\ref{fig:high_tc}.)
We will show that the line nodes are stable.
We perturb the parameters of the Hamiltonian as
\begin{eqnarray}
&&\epsilon\rightarrow \epsilon+\delta\epsilon,
\quad
\psi\rightarrow \psi+\delta\psi,
\nonumber\\
&&{\bm g}\equiv 0\rightarrow{\bm g}=\delta{\bm g},
\quad 
{\bm d}\equiv 0\rightarrow{\bm d}=\delta{\bm d}. 
\end{eqnarray}
First, consider the case of $\delta{\bm g}=\delta{\bm d}=0$.
In this case, the perturbation does not break the parity symmetry.
Now Eqs.(\ref{eq:node1}) and (\ref{eq:node2}) reduce to
\begin{eqnarray}
\epsilon({\bm k})+\delta\epsilon({\bm k})=0,
\quad
\psi({\bm k})+\delta\psi({\bm k})=0. 
\label{eq:node_high_tc}
\end{eqnarray}
These equations can be satisfied by a small change of the
position of the line nodes ${\bm k}={\bm k}_0\rightarrow {\bm k}={\bm
k}_0+\delta{\bm k}_0$.
Indeed, expanding Eq.(\ref{eq:node_high_tc}) around
${\bm k}={\bm k}_0$, we have
\begin{eqnarray}
&&\nabla\epsilon({\bm k}_0)\cdot\delta{\bm k}_0+\delta\epsilon({\bm k}_0)=0,
\nonumber\\ 
&&\nabla\psi({\bm k}_0)\cdot\delta{\bm k}_0+\delta\psi({\bm k}_0)=0.
\end{eqnarray}
(Here we have used $\epsilon({\bm k}_0)=\psi({\bm k}_0)=0$.)
For any $\delta\epsilon$ and $\delta\psi$, there exists $\delta{\bm k}_0$
which satisfies these equations. 
Therefore, the perturbation moves the position of the line nodes, but it
does not remove the line nodes.

\begin{figure}[h]
\begin{center}
\includegraphics[width=9cm]{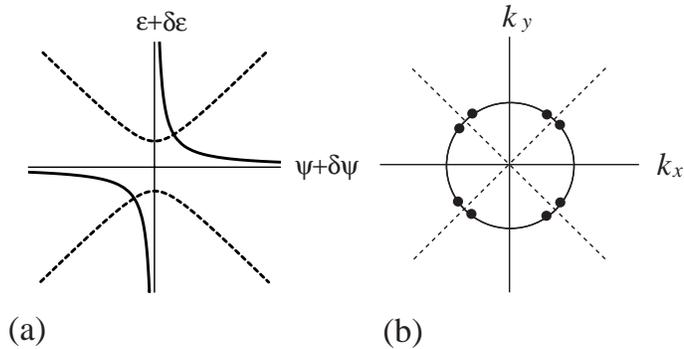}
\caption{(a) Two curves given by Eqs.(\ref{eq:node1_high_tc}) (the solid
 lines) and (\ref{eq:node2_high_tc}) (the dotted lines). Here we show the
 case of $\delta{\bm g}^2-\delta{\bm d}^2>0$ and $\delta {\bm g}\cdot\delta{\bm d}>0$. 
(b) A schematic figure of the line nodes in high-$T_{\rm c}$
 superconductors with a
 parity breaking perturbation.}
\label{fig:high_tc_2}
\end{center}
\end{figure}

When $\delta{\bm g}$ and $\delta{\bm d}$ are not zero,
Eqs.(\ref{eq:node1}) and (\ref{eq:node2}) become
\begin{eqnarray}
&&(\epsilon+\delta\epsilon)\cdot(\psi+\delta\psi)
=\delta{\bm g}\cdot\delta{\bm d}.  
\label{eq:node1_high_tc}
\\
&&(\epsilon+\delta\epsilon)^2-(\psi+\delta\psi)^2
=\delta{\bm g}^2-\delta{\bm d}^2,
\label{eq:node2_high_tc}
\end{eqnarray}
These two equations give two curves in the
$(\psi+\delta\psi, \epsilon+\delta\epsilon)$ plane.
As is shown in Fig.\ref{fig:high_tc_2}(a),
these two curves always have two intersection points near the origin. 
Therefore, we have two different conditions of line nodes corresponding
to the two intersection points.
This implies that each line node splits into two after the perturbation
as illustrated in Fig.\ref{fig:high_tc_2}(b).
Although the number of line nodes becomes eight, 
the line nodes do not vanish by the perturbation. 

These line nodes vanish if the time-reversal symmetry is broken.
For example, by deforming $\psi({\bm k})$ as
\begin{eqnarray}
\psi({\bm k})\rightarrow \psi({\bm k})+i\eta, 
\quad
(\eta={\rm const.}),
\end{eqnarray}
we can completely remove the line nodes.

\subsection{polar state in $p$-wave paring}
\label{sec:ex_polar}

\begin{figure}[h]
\begin{center}
\includegraphics[width=5cm]{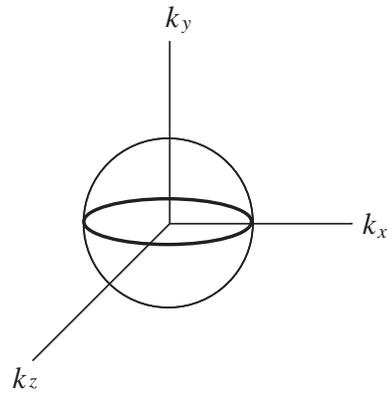}
\caption{A line node of the polar state in $p$-wave paring.}
\label{fig:polar}
\end{center}
\end{figure}

Here we consider the polar state in  $p$-wave paring.
The polar state is given by
\begin{eqnarray}
\epsilon({\bm k})=\frac{{\bm k}^2}{2m}-\mu, 
\quad
{\bm d}({\bm k})=\Delta_0\hat{{\bm z}}k_z, 
\end{eqnarray}
and
\begin{eqnarray}
\psi({\bm k})={\bm g}({\bm k})=0. 
\end{eqnarray}
It is a solution of the Ginzburg-Landau theory neglecting the spin-orbit
coupling \cite{Leggett}.
Obviously, the gap has a line node on the equator. See Fig.\ref{fig:polar}.

The line node is unstable, because, for example,
the following deformation of the gap function removes the line node completely,
\begin{eqnarray}
{\bm d}({\bm k})\rightarrow {\bm d}({\bm k})+\eta
(\hat{{\bm x}}k_x+\hat{\bm y}k_z), 
\end{eqnarray}
where $\eta$ is a small real number.
This line node is not given by an intersection line of
Eqs.(\ref{eq:node1}) and (\ref{eq:node2}), thus 
it is accidental and topologically unstable.

\subsection{mixed singlet-triplet states}
\label{sec:ex_mixed}

For noncentrosymmetric materials, the parity symmetry is broken in the
normal states.
In the presence of the spin-orbit interaction, the absence of inversion
symmetry give rise to parity breaking terms ${\bm g}({\bm k})$ in the
normal states. 
In such systems, singlet and triplet parings can be mixed in the
superconducting states \cite{GR}.  
We examine stability of line nodes in the mixed singlet-triplet
superconducting states.

\begin{figure}[h]
\begin{center}
\includegraphics[width=5cm]{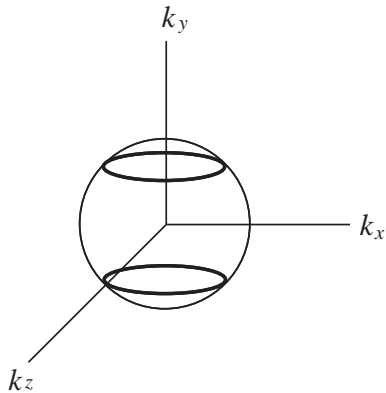}
\caption{A schematic figure of line nodes on the Fermi surface of
 ${\rm CePt_3Si}$ \cite{HWFS}. The sphere and the solid lines denote the
 Fermi surface and the line nodes, respectively.}
\label{fig:cept3si}
\end{center}
\end{figure}

As a concrete example, consider a $s+p$ superconducting state proposed to
explain the superconducting state in ${\rm CePt_3Si}$.
In this material, the existence of line nodes was reported experimentally
\cite{IKMBYSO}.
In ${\rm CePt_3Si}$, the absence of inversion symmetry give rise to the
Rashba interaction, 
\begin{eqnarray}
{\bm g}({\bm k})=\alpha\sqrt{\frac{3}{2}}(-\hat{\bm x}k_y+\hat{\bm y}k_x),
\end{eqnarray}
where $\alpha$ is a real number.
The following gap function was proposed to explain the line node in the
superconducting state \cite{HWFS},
\begin{eqnarray}
\psi({\bm k})=\Psi_0,
\quad
{\bm d}({\bm k})=\Delta_0(-\hat{\bm x}k_y+\hat{\bm y}k_x).
\end{eqnarray}
Here $\Psi_0$ and $\Delta_0$ are real numbers.
As is shown in Ref.\cite{HWFS}, 
one can show that 
under a suitable choice of $\alpha$, $\Psi_0$ and $\Delta_0$, 
$E_{-}({\bm k})$ has two line nodes. 
($E_{+}({\bm k})$ has no node.)
We show a schematic picture of the nodes in
Fig.\ref{fig:cept3si}.
As in the high-$T_{\rm c}$ case, it can be shown that the line nodes are stable
against any small time-reversal invariant perturbations;
The line nodes are intersection lines of the surfaces 
given by Eqs.(\ref{eq:node1}) and ({\ref{eq:node2}}) in momentum space. 
Under a small perturbation,
\begin{eqnarray}
&&\epsilon\rightarrow \epsilon+\delta\epsilon,
\quad
\psi\rightarrow \psi+\delta\psi, 
\nonumber\\
&&{\bm g}\rightarrow {\bm g}+\delta{\bm g},
\quad
{\bm d}\rightarrow{\bm d}+\delta{\bm d},
\end{eqnarray}
these surfaces moves slightly, but the intersection lines do not vanish unless
the lines shrink to points.

In a similar manner, we can show generally that 
line nodes are topologically stable in mixed singlet-triplet
superconducting states with time-reversal invariance. 
For example, the line nodes proposed for ${\rm Li_2Pd_3B}$ and
${\rm Li_2Pt_3B}$ \cite{YAHBVTSS} are topologically stable.

\section{$Z_2$ Topological Number}
\label{sec:topological_number}

In the previous sections, we have seen that line nodes can be stable in
superconducting states with time-reversal invariance.
In this section, we show that the stability is explained by the existence
of topological numbers defined by wavefunctions of quasiparticles.
The topological numbers are closely related to a real
structure of the Hamiltonian $H({\bm k})$.
Although $H({\bm k})$ is not a real matrix in general,
it has another real structure due to the time-reversal symmetry.

\subsection{Real structure and ${\bm Z}_2$ topological number}

\subsubsection{Real structure}

In order to find a real structure, it is convenient to consider the following
$8\times8$ Hamiltonian,
\begin{eqnarray}
{\cal H}({\bm k})=
\left(
\begin{array}{cc}
H({\bm k}) & 0\\
0 & H(-{\bm k})
\end{array}
\right), 
\end{eqnarray}
and its eigenvalue equation,
\begin{eqnarray}
{\cal H}({\bm k})\Psi({\bm k})=E({\bm k})\Psi({\bm k}). 
\end{eqnarray}
The eigenvalue $E({\bm k})$ is the same as that of $H({\bm k})$.

Using the time-reversal symmetry of $H({\bm k})$, we can show 
\begin{eqnarray}
\Xi{\cal H}({\bm k})\Xi^{-1}= {\cal H}({\bm k})^{*},
\label{eq:real}
\end{eqnarray}
where $\Xi$ is defined by
\begin{eqnarray}
\Xi=
\left(
\begin{array}{cc}
0 &\Theta\\
-\Theta &0
\end{array}
\right), 
\quad \Xi^2=1. 
\end{eqnarray}
The relation (\ref{eq:real}) implies a real structure of ${\cal H}({\bm
k})$.
To see this, we introduce an operator ${\cal K}=\Xi K$ where $K$ is the
ordinary complex-conjugation operator.
The new operator obeys ${\cal K}^2=1$, then we can think of
it as a new complex-conjugation operator.
The extended Hamiltonian ${\cal H}({\bm k})$ is ``real'' in terms of the
new complex-conjugate operation, 
\begin{eqnarray}
{\cal K}{\cal H}({\bm k}){\cal K}^{-1}={\cal H}({\bm k}). 
\end{eqnarray}

Now we can impose the reality condition on the eigenfunction
$\Psi({\bm k})$,
\begin{eqnarray}
\Psi({\bm k})={\cal K}\Psi({\bm k})(\equiv \Xi\Psi({\bm k})^{*}).
\label{eq:reality}
\end{eqnarray}
Using an eigenfunction of $H({\bm k})$, 
\begin{eqnarray}
H({\bm k})u({\bm k})=E({\bm k})u({\bm k}), 
\end{eqnarray}
we find the following two independent eigenfunctions of ${\cal H}({\bm
k})$ which satisfy the reality condition,
\begin{eqnarray}
\Psi_{\rm R}({\bm k})=\frac{1}{2}
\left(
\begin{array}{c}
u({\bm k}) \\
-\Theta u({\bm k})^{*}
\end{array}
\right), 
\end{eqnarray}
and
\begin{eqnarray}
\Psi_{\rm I}({\bm k})=-\frac{i}{2}
\left(
\begin{array}{c}
u({\bm k}) \\
\Theta u({\bm k})^{*}
\end{array}
\right).
\end{eqnarray}
Since $u({\bm k})$ is rewritten as 
\begin{eqnarray}
\left(
\begin{array}{c}
 u({\bm k})\\
0
\end{array}
\right) 
=\Psi_{\rm R}({\bm k})+i\Psi_{\rm I}({\bm k}),
\end{eqnarray}
these two wavefunctions correspond to the ``real'' and
``imaginary'' parts of $u({\bm k})$, respectively.
If $u({\bm k})$ is normalized as
\begin{eqnarray}
u({\bm k})^{\dagger}u({\bm k})=1,  
\end{eqnarray}
the normalization of $\Psi_A({\bm k})$ is given by
\begin{eqnarray}
\Psi_A({\bm k})^{\dagger}\Psi_B({\bm k})=\frac{1}{2}\delta_{AB},
\label{eq:normalization}
\end{eqnarray}
where $A,B={\rm R}, {\rm I}$.

\subsubsection{${\bm Z}_2$ topological number}

\begin{figure}[h]
\begin{center}
\includegraphics[width=6cm]{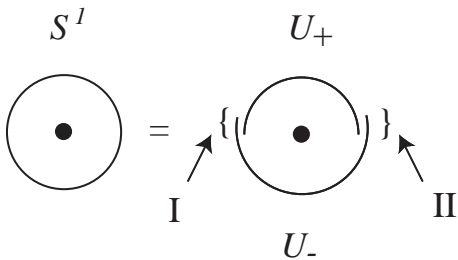}
\caption{$S^1$ around a line node. $S^1$ is covered by $U_+$ and $U_-$.
The overlap of $U_+$ and $U_-$ consists of the regions I and II.}
\label{fig:node}
\end{center}
\end{figure}

To define topological numbers of a line node, let us 
consider first an infinitesimal circle $S^{1}$ around the line node and
solve the eigenequation $H({\bm k})u({\bm k})=E({\bm k})u({\bm k})$ on $S^1$. 
(See Fig.\ref{fig:node}.)
Because of the gauge freedom 
\begin{eqnarray}
u({\bm k})\rightarrow e^{i\theta({\bm k})}u({\bm k}),
\end{eqnarray}
$u({\bm k})$ is not determined uniquely. 
We have to fix the gauge.
Generally, a gauge fixed solution has a singularity on $S^1$, thus it
can not be well-defined on the entire $S^1$.  
Two or more solutions with different gauge fixings are needed to cover
the entire $S^1$.
We consider a couple of solutions $u^{(1)}({\bm k})$ and $u^{(2)}({\bm
k})$, and demand that the first component of
$u^{(1)}({\bm k})$ and the second one of $u^{(2)}({\bm k})$ are real.
Because of these different demands, the solutions $u^{(1)}({\bm k})$
and $u^{(2)}({\bm k})$ have different singularities from each other.
Therefore, we can cover the entire $S^1$ by using them.

Let us next consider the Kramers doublet of $u^{(2)}({\bm k})$,
$\Theta u^{(2)}(-{\bm k})^{*}$. 
By definition, the Kramers doublet has the real first component as well
as $u^{(1)}({\bm k})$.
Therefore the following two possibility arises, 
\begin{enumerate}
 \item $\Theta u^{(2)}(-{\bm k})^{*}=u^{(1)}({\bm k})$ (or
       $-u^{(1)}({\bm k})$),
 \item $\Theta u^{(2)}(-{\bm k})^{*}\neq \pm u^{(1)}({\bm k})$.
\end{enumerate}
Here we concentrate on the latter case,
since only this is relevant to the topological numbers of the line node 
\footnote{We found that the former case arises around a degenerate point of
$E_{+}({\bm k})=E_{-}({\bm k})$. This is related to the topological
number of the degenerate point which will be mentioned in
Sec.\ref{sec:conclusion}.}.
In this case, $\Theta u^{(2)}(-{\bm k})^{*}$ has
a different singularity from $u^{(1)}({\bm k})$, thus we can use $\Theta
u^{(2)}(-{\bm k})^{*}$ instead of $u^{(2)}({\bm k})$ to cover the entire
region of $S^1$.
After all,
the following two solutions with the real first component are obtained,
\begin{eqnarray}
u^{(+)}({\bm k})\equiv u^{(1)}({\bm k}),
\quad
u^{(-)}({\bm k})\equiv \Theta u^{(2)}(-{\bm k})^{*}.
\end{eqnarray}
If the solutions $u^{(+)}({\bm k})$ and $u^{(-)}({\bm k})$ are nonsingular
on $U_{+}$ and $U_{-}$, respectively,
$S^1$ is given by the union of $U_{+}$ and $U_{-}$,
\begin{eqnarray}
S^1=U_{+}\cup U_{-}. 
\end{eqnarray}

Now it is convenient to introduce the eigenfunctions of ${\cal H}({\bm k})$
instead of $u^{(\sigma)}({\bm k})$ $(\sigma=\pm)$.
The eigenfunctions of ${\cal H}({\bm k})$ with the reality condition
(\ref{eq:reality}) are constructed from $u^{(\sigma)}({\bm k})$ as
\begin{eqnarray}
&&\Psi^{(\sigma)}_{\rm R}({\bm k})=\frac{1}{2}
\left(
\begin{array}{c}
u^{(\sigma)}({\bm k}) \\
-\Theta u^{(\sigma)}({\bm k})^{*}
\end{array}
\right), 
\nonumber\\
&&\Psi^{(\sigma)}_{\rm I}({\bm k})=-\frac{i}{2}
\left(
\begin{array}{c}
u^{(\sigma)}({\bm k}) \\
\Theta u^{(\sigma)}({\bm k})^{*}
\end{array}
\right). 
\end{eqnarray}
It is easily find that the eigenfunctions $\Psi_{A}^{(\sigma)}({\bm
k})$'s $(A={\rm R}, {\rm I})$ are regular in $U_{\sigma}$. 

On the overlap $U_{+}\cap U_{-}$, both $\Psi^{(+)}_{A}({\bm k})$ and
$\Psi^{(-)}_{B}({\bm k})$ are regular and they are related by the
transition function $T_{AB}({\bm k})$,
\begin{eqnarray}
\Psi^{(+)}_A({\bm k})=T_{AB}({\bm k})\Psi^{(-)}_B({\bm k}).
\end{eqnarray}
The reality condition (\ref{eq:reality}) and the normalization
(\ref{eq:normalization}) imply that the
transition function $T({\bm k})(\equiv T_{AB}({\bm k}))$ is an element
of ${\rm O}(2)$. 
Moreover, $T({\bm k})$ reduces to 
\begin{eqnarray}
T({\bm k})=
\left(
\begin{array}{cc}
t_{\rm R}({\bm k}) & 0\\
0 & t_{\rm I}({\bm k})
\end{array}
\right),
\quad 
t_{\rm R}({\bm k})=t_{\rm I}({\bm k})=\pm 1, 
\end{eqnarray}
since $\Psi^{(\sigma)}_{\rm R}({\bm k})$'s have the real first
components but $\Psi^{(\sigma)}_{\rm I}({\bm k})$'s have the imaginary
ones.
Therefore, $\Psi_{\rm R}^{(+)}({\bm k})$ (or $\Psi_{\rm I}^{(+)}({\bm k})$)
is identified with $\Psi_{\rm R}^{(-)}({\bm k})$ ($\Psi_{\rm
I}^{(-)}({\bm k})$) on $U_+\cap U_-$.
As we will show immediately, the transition function determines the
global topology of the wavefunctions.

For the sake of simplicity, we assume in the following that both $U_{+}$ and
$U_{-}$ are connected regions as illustrated in Fig.\ref{fig:node}.
In this case, the overlap $U_{+}\cap U_{-}$ consists of two regions I and II.
The generalization to other cases in which $U_{+}$ and $U_{-}$ consist of many
disconnected segments is straightforward.

\begin{figure}[h]
\begin{center}
\includegraphics[width=7.5cm]{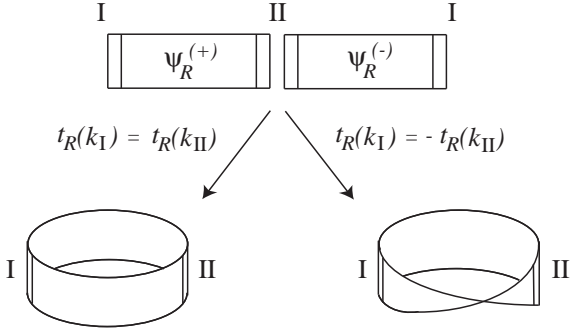}
\caption{Possible configurations around a line node. (a) A
 contractable loop. (b) A noncontractable loop.}
\label{fig:mobius}
\end{center}
\end{figure}

Let us examine the topology of $\Psi^{(\sigma)}_{\rm R}({\bm k})$. 
On $S^1$, it has two transition functions $t_{\rm R}({\bm k_{\rm I}})$ and
$t_{\rm R}({\bm k}_{\rm II})$ where ${\bm k}_{\rm I}$ and ${\bm k}_{\rm
II}$ are arbitrary points on the regions I and II, respectively.
If $t_{\rm R}({\bm k}_{\rm I})=t_{\rm R}({\bm k}_{\rm II})$, 
$\Psi^{(+)}_{\rm R}({\bm k})$ are glued to $\Psi^{(-)}_{\rm R}({\bm k})$
without twisting, thus we have a trivial topology similar to a cylinder.
(See Fig \ref{fig:mobius}.) 
This configuration is a contractable loop in the
Hilbert space, therefore nothing interesting happens. 
On the other hand, if $t_{\rm R}({\bm k}_{\rm I})=-t_{\rm R}({\bm
k}_{\rm II})$, $\Psi^{(+)}_{\rm R}({\bm k})$ and $\Psi^{(-)}_{\rm
R}({\bm k})$ are glued together with twisting, thus we have a
nontrivial topology similar to the M\"{o}bius strip \cite{EGH}.
Since this configuration is a noncontractable loop around the line node,
the line node can be considered as a kind of topological defect (vortex).
Therefore, in this case the line node is stable against a continuous
deformation of the Hamiltonian.
If we twist the wavefunction in twice, we have a trivial configuration again. 
Thus, the corresponding homotopy group is $\pi_1={\bm Z}_2$.

By this argument, we naturally introduce the following ${\bm Z}_2$
topological number $I_{\rm R}$,
\begin{eqnarray}
I_{\rm R}
=\frac{\ln 
\left(t_{\rm R}({\bm k}_{\rm I})t_{\rm R}({\bm k}_{\rm II})\right)
}
{\pi i} \quad ({\rm mod.}2).
\end{eqnarray}
It takes
\begin{eqnarray}
I_{\rm R}=
\left\{
\begin{array}{ll}
0, & \mbox{for $t_{\rm R}({\bm k}_{\rm I})=t_{\rm R}({\bm k}_{\rm II})$},\\
1, & \mbox{for $t_{\rm R}({\bm k}_{\rm I})=-t_{\rm R}({\bm k}_{\rm II})$}.
\end{array} 
\right.
\end{eqnarray}
The line node is topologically stable when $I_{\rm R}=1$ $({\rm
mod.}2)$, and the stability is explained by the conservation of the
topological number.

In a similar manner, another
${\bm Z}_2$ topological number can be defined as, 
\begin{eqnarray}
I_{\rm I}
=\frac{\ln 
\left(t_{\rm I}({\bm k}_{\rm I})t_{\rm I}({\bm k}_{\rm II})\right)
}
{\pi i}\quad ({\rm mod.}\,2).   
\end{eqnarray}
Since $t_{\rm I}({\bm k})$ is equal to $t_{\rm R}({\bm k})$, it is
easily seen that $I_{\rm I}=I_{\rm R}$. 

Unless the time-reversal symmetry is broken, 
the two topological numbers $I_{\rm R}$ and $I_{\rm I}$ are conserved
independently.
However, once the time-reversal invariance is lost,
they are not conserved.
The reality of the Hamiltonian (\ref{eq:reality}) is lost and  
$\Psi_{\rm R}({\bm k})$ cannot be distinguished from $\Psi_{\rm I}({\bm
k})$. 
Although the summation of $I_{\rm R}$ and $I_{\rm I}$ may be conserved,
it is identically zero
\begin{eqnarray}
I_{\rm R}+I_{\rm I}=0 \quad({\rm mod.}2).
\end{eqnarray}
Therefore, the line node loses the topological stability and it
disappears in general.

It can be shown that the topological numbers are gauge invariant if
the gauge transformation is nonsingular.
Consider the following gauge transformation 
\begin{eqnarray}
u({\bm k})\rightarrow u'({\bm k})=e^{i\theta({\bm k})}u({\bm k}), 
\label{eq:nonsingular}
\end{eqnarray}
where $\theta({\bm k})$ is a nonsingular function of ${\bm k}$.
When $\theta({\bm k})$ is nonsingular, the regions $U_{+}$ and $U_{-}$
in which $u^{(+)}{}'({\bm k})$ and $u^{(-)}{}'({\bm k})$ are nonsingular
remain the same as before.
The $U(1)$ gauge transformation (\ref{eq:nonsingular}) induces the
following ${\rm SO}(2)$ gauge transformation on $\Psi^{(\sigma)}_A({\bm k})$, 
\begin{eqnarray}
\Psi^{(\sigma)}_A{}'({\bm k})=O_{AB}({\bm k})\Psi^{(\sigma)}_B({\bm k}),
\end{eqnarray}
where
\begin{eqnarray}
O({\bm k})\equiv O_{AB}({\bm k})=\left(
\begin{array}{cc}
\cos\theta({\bm k}) &-\sin\theta({\bm k}) \\
\sin\theta({\bm k}) & \cos\theta({\bm k})
\end{array}
\right). 
\end{eqnarray}
The new transformation function $T'({\bm k})$ between
$\Psi^{(\sigma)}_A{}'({\bm k})$'s is given by
\begin{eqnarray}
T'({\bm k})=O({\bm k})T({\bm k})O({\bm k})^{\rm T}. 
\end{eqnarray}
Because the ${\rm O}(2)$ matrices $O({\bm k})$ and $T({\bm k})$ commute
with each other, we obtain
\begin{eqnarray}
T'({\bm k})=T({\bm k}). 
\end{eqnarray}
Therefore, the topological numbers $I_{\rm R}$ and $I_{\rm I}$
defined by $T({\bm k})$ are gauge invariant.

\subsection{high-${\rm T_c}$ materials}
\label{sec:topological_number_high_tc}

\begin{figure}[h]
\begin{center}
\includegraphics[width=6cm]{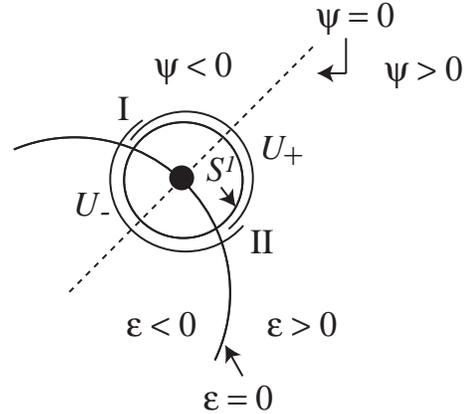}
\caption{$S^1$ around a line node in high-$T_{\rm c}$ superconductors}
\label{fig:s1_high_tc}
\end{center}
\end{figure}

The Hamiltonian of high-$T_{\rm c}$ superconductors is
\begin{eqnarray}
H({\bm k})=
\left(
\begin{array}{cc}
\epsilon({\bm k}) & i\psi({\bm k})\sigma_2\\
-i\psi({\bm k})\sigma_2 & -\epsilon({\bm k})
\end{array}
\right).
\end{eqnarray}
This superconducting state is unitary, thus the quasiparticle energies
$E_{\pm}({\bm k})$ are degenerate in the entire momentum space, 
\begin{eqnarray}
E_{\pm}({\bm k})=E({\bm k})=\sqrt{\epsilon({\bm k})^2+\psi({\bm k})^2}. 
\end{eqnarray}
Since the parity is conserved and $H({\bm k})$ is real,
$H({\bm k})$ commutates with $\Theta$,  
\begin{eqnarray}
\Theta H({\bm k})=H({\bm k})\Theta. 
\end{eqnarray}
Therefore, the eigenfunctions $u_i({\bm k})$
($i=1,2$) with $E({\bm k})$ can be the eigenfunction of $\Theta$
simultaneously,
\begin{eqnarray}
\Theta u_1({\bm k})=iu_1({\bm k}),
\quad
\Theta u_2({\bm k})=-iu_2({\bm k}).
\end{eqnarray}
Consider a line node shown in Fig.\ref{fig:s1_high_tc}.
The eigenfunctions regular in $U_{+}$ are given by 
\begin{eqnarray}
&&u_1^{(+)}({\bm k})=
\frac{1}{2\sqrt{E(E+\epsilon)}}
\left(
\begin{array}{c}
(E+\epsilon) \\
i(E+\epsilon)\\
-i\psi\\
\psi
\end{array}
\right), 
\nonumber\\
&&u_2^{(+)}({\bm k})=
\frac{1}{2\sqrt{E(E+\epsilon)}}
\left(
\begin{array}{c}
E+\epsilon \\
-i(E+\epsilon)\\
i\psi\\
\psi
\end{array}
\right),
\end{eqnarray}
and those regular in $U_{-}$ are
\begin{eqnarray}
&&u_1^{(-)}({\bm k})= 
\frac{1}{2\sqrt{E(E-\epsilon)}}
\left(
\begin{array}{c}
\psi\\
i\psi\\
-i(E-\epsilon)\\
E-\epsilon
\end{array}
\right), 
\nonumber\\
&&u_2^{(-)}({\bm k})=
\frac{1}{2\sqrt{E(E-\epsilon)}}
\left(
\begin{array}{c}
\psi \\
-i\psi\\
i(E-\epsilon)\\
E-\epsilon
\end{array}
\right). 
\end{eqnarray}
In a similar manner as $t_{A}({\bm k})$ and $I_{\rm A}$ $(A={\rm R},{\rm
I})$ in the previous subsection, 
the transition functions $t_{a A}({\bm k})$ and the topological number $I_{aA}$ $(a=1,2,\,A={\rm R},{\rm
I})$ are constructed from $u_a^{(\sigma)}({\bm k})$ $(a=1,2,\,\sigma=\pm)$. 
The transition functions become
\begin{eqnarray}
&&t_{1{\rm R}}({\bm k})=t_{1{\rm I}}({\bm k})={\rm sgn}\psi({\bm k}), 
\nonumber\\
&&t_{2{\rm R}}({\bm k})=t_{2{\rm I}}({\bm k})={\rm sgn}\psi({\bm k}).
\end{eqnarray}
Because the transition functions have different signatures between regions
I and II in Fig.\ref{fig:s1_high_tc}, 
the topological numbers calculated on $S^1$ are given by
\begin{eqnarray}
I_{1{\rm R}}=I_{1{\rm I}}=1, 
\quad
I_{2{\rm R}}=I_{2{\rm I}}=1. 
\end{eqnarray}
As long as the parity is conserved, the eigenfunctions of $H({\bm k})$ stay 
the eigenfunctions of $\Theta$.
Therefore, $I_{1A}$ and $I_{2A}$ $(A={\rm R},{\rm I})$ are not mixed,
and the stability of the line node is ensured by conservation of
these topological numbers.

If the parity is broken, the degeneracy between
$E_{+}({\bm k})$ and $E_{-}({\bm k})$ is resolved and only $E_{-}({\bm
k})$ has line nodes. 
When the perturbation is small enough, the topological
numbers $I_{\rm R}$ and $I_{\rm I}$ defined by the eigenfunction with
$E_{-}({\bm k})$ are calculated on the same $S^1$ as
\begin{eqnarray}
I_{\rm R}=I_{1{\rm R}}+I_{2{\rm R}}=0,
\quad
I_{\rm I}=I_{1{\rm I}}+I_{2{\rm I}}=0.
\label{eq:high_tc_split}
\end{eqnarray}
This is consistent with the splitting of the line nodes shown in
Sec.\ref{sec:ex_high_tc}; The line node in Fig.\ref{fig:s1_high_tc}
is divided into two as is shown in Sec.\ref{sec:ex_high_tc}, and each
divided line node has the topological numbers $I_{\rm R}=I_{\rm I}=1$.
Because $S^1$ encloses both the divided line nodes, the topological numbers
calculated on $S^1$ become $I_{\rm R}=I_{\rm I}=1+1=0$ which is the same
as Eq.(\ref{eq:high_tc_split}).

\subsection{polar state in $p$-wave paring}

\begin{figure}[h]
\begin{center}
\includegraphics[width=3.5cm]{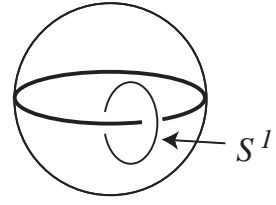}
\caption{$S^1$ around the line node in the polar state.}
\label{fig:s1_polar}
\end{center}
\end{figure}

The Hamiltonian of the polar state in $p$-wave paring is 
\begin{eqnarray}
H({\bm k})=
\left(
\begin{array}{cc}
\epsilon({\bm k}) & d_{3}({\bm k})\sigma_1 \\
d_{3}({\bm k})\sigma_1 &-\epsilon({\bm k})
\end{array}
\right). 
\end{eqnarray}
The quasiparticle energies $E_{\pm}({\bm k})=E({\bm k})$ are given by  
\begin{eqnarray}
E({\bm k})=\sqrt{\epsilon({\bm k})^2+d_3({\bm k})^2}. 
\end{eqnarray}
Consider a circle $S^1$ around the line node in the equator.
(See Fig.\ref{fig:s1_polar}.)
The circle $S^1$ is covered by the following solutions,
\begin{eqnarray}
&&u_1^{(+)}({\bm k})= 
\frac{1}{2\sqrt{E(E+\epsilon)}}
\left(
\begin{array}{c}
E+\epsilon \\
E+\epsilon \\
d_3 \\
d_3
\end{array}
\right), 
\nonumber\\
&& u_1^{(-)}({\bm k})= 
\frac{1}{2\sqrt{E(E-\epsilon)}}
\left(
\begin{array}{c}
d_3 \\
d_3 \\
E-\epsilon\\
E-\epsilon
\end{array}
\right), 
\nonumber\\
&&u_2^{(+)}({\bm k})= 
\frac{1}{2\sqrt{E(E+\epsilon)}}
\left(
\begin{array}{c}
E+\epsilon \\
-(E+\epsilon) \\
-d_3 \\
d_3
\end{array}
\right), 
\nonumber\\
&&u_2^{(-)}({\bm k})= 
\frac{1}{2\sqrt{E(E-\epsilon)}}
\left(
\begin{array}{c}
-d_3 \\
d_3 \\
E-\epsilon \\
-(E-\epsilon)
\end{array}
\right). 
\end{eqnarray}
Using these solutions, we obtain the transition functions as follows,
\begin{eqnarray}
&&t_{1{\rm R}}({\bm k})=t_{1{\rm I}}({\bm k})={\rm sgn}d_3({\bm k}),
\nonumber\\
&&t_{2{\rm R}}({\bm k})=t_{2{\rm I}}({\bm k})=-{\rm sgn}d_3({\bm k}). 
\end{eqnarray}
From this, the topological numbers are calculated as 
\begin{eqnarray}
I_{1{\rm R}}=I_{1{\rm I}}=1, 
\quad
I_{1{\rm R}}=I_{1{\rm I}}=1.
\end{eqnarray}
At a first glass, it seems to suggest that the line node has nontrivial
topological numbers.
However, on the contrary to the high-$T_{\rm c}$ case, there is no
quantum number which distinguishes $u_{1}^{(\sigma)}({\bm k})$ from
$u_{2}^{(\sigma)}({\bm k})$ $(\sigma=\pm)$.  
Therefore, only the summations $I_{1A}+I_{2A}=0$ $(A={\rm R},{\rm I})$
are preserved, and the line node has no nontrivial topological numbers.
This result is consistent with the result in Sec.\ref{sec:ex_polar};
The line node is unstable against a small perturbation of
$H({\bm k})$.

\subsection{mixed singlet-triplet state}

The Hamiltonian of mixed singlet-triplet states is  
\begin{eqnarray}
H({\bm k})=
\left(
\begin{array}{cc}
\epsilon({\bm k})+{\bm g}({\bm k})\cdot{\bm \sigma} & \Delta({\bm k})\\
\Delta({\bm k})^{\dagger} & -\epsilon({\bm k})
+{\bm g}({\bm k})\cdot{\bm \sigma}^{*}
\end{array}
\right), 
\end{eqnarray}
where
\begin{eqnarray}
\Delta({\bm k})
=i\psi({\bm k})\sigma_2+i{\bm d}({\bm k})\cdot{\bm \sigma}\sigma_2.
\end{eqnarray}
Here $\psi({\bm k})$ and ${\bm d}({\bm k})$ are real functions.
As was shown in Ref.\cite{FAKS}, ${\bm d}({\bm k})$ is
proportional to ${\bm g}({\bm k})$ in mixed singlet-triplet states, 
\begin{eqnarray}
{\bm d}({\bm k})=c{\bm g}({\bm k}).
\end{eqnarray}
For the sake of simplicity, we assume $c>0$ in the following.
$E_{-}({\bm k})$ is given by
\begin{eqnarray}
&&E_{-}({\bm k})
\nonumber\\
&=&\sqrt{(\epsilon({\bm k})-{\rm sgn}\psi({\bm k})|{\bm g}({\bm k})|)^2+
(|\psi({\bm k})|-|{\bm d}({\bm k})|)^2}. 
\nonumber\\
\end{eqnarray}
The eigenfunction has the following two forms,
\begin{eqnarray}
u^{(+)}({\bm k})
&=&\frac{1}{\sqrt{2E_{-}(E_{-}+\epsilon-{\rm sgn}\psi|{\bm g}|)}} 
\nonumber\\
&\times&
\left(
\begin{array}{c}
\left(E_{-}+\epsilon-{\rm sgn}\psi|{\bm g}|\right)\phi\\
-i{\rm sgn}\psi\left(|\psi|-|{\bm d}|\right)
\sigma_2\phi
\end{array}
\right),
\nonumber\\
u^{(-)}({\bm k})
&=&\frac{1}{\sqrt{2E_{-}(E_{-}-\epsilon+{\rm sgn}\psi|{\bm g}|)}} 
\nonumber\\
&\times&
\left(
\begin{array}{c}
{\rm sgn}\psi\left(|\psi|-|{\bm d}|\right)\phi\\
-i\left(E_{-}-\epsilon+{\rm sgn}\psi|{\bm g}|\right)\sigma_2\phi
\end{array}
\right),
\end{eqnarray}
where $\phi({\bm k})$ satisfies
\begin{eqnarray}
{\bm d}({\bm k})\cdot{\bm \sigma}\phi({\bm k})
=-{\rm sgn}\psi({\bm k})|{\bm d}({\bm k})|\phi({\bm k}). 
\end{eqnarray}
The first components of $u^{(\pm)}({\bm k})$ are real if we choose 
\begin{eqnarray}
&&\phi({\bm k})=
\nonumber\\
&&\left\{
\begin{array}{cc}
\frac{1}{\sqrt{2|{\bm d}|(|{\bm d}|-d_3)}}
\left(
\begin{array}{c}
|{\bm d}|-d_3 \\
-d_1-id_2
\end{array}
\right),& 
(\mbox{for ${\rm sgn}\psi>0$})
 \\
\frac{1}{\sqrt{2|{\bm d}|(|{\bm d}|+d_3)}}
\left(
\begin{array}{c}
|{\bm d}|+d_3 \\
d_1+id_2
\end{array}
\right),&
(\mbox{for ${\rm sgn}\psi<0$}) 
\end{array}
\right. .
\nonumber\\
\end{eqnarray}
The transition functions obtained from $u^{(+)}({\bm k})$ and
$u^{(-)}({\bm k})$ are
\begin{eqnarray}
t_{\rm R}({\bm k})=t_{\rm I}({\bm k})
={\rm sgn}\psi({\bm k})\cdot{\rm sgn}(|\psi({\bm k})|-|{\bm d}({\bm k})|). 
\end{eqnarray}
From these transition functions, we can calculate the topological numbers
$I_{\rm R}$ and $I_{\rm I}$ of line nodes in mixed singlet-triplet
superconducting states. 
For example, the straightforward calculation shows that both the line
nodes of ${\rm CeSi_3Pt}$ in Fig.\ref{fig:cept3si} have nontrivial topological
numbers, $I_{\rm R}=I_{\rm I}=1$. 
We can also show that the line nodes proposed for ${\rm Li_2Pd_3B}$ and
${\rm Li_2Pt_3B}$ \cite{YAHBVTSS} have nontrivial topological numbers.
These results are consistent with the stability of the line nodes
examined in Sec.\ref{sec:ex_mixed}.

Note that the time-reversal symmetry is essential for the stability of the line
nodes.
For example, when $\psi({\bm k})$ becomes complex, the time-reversal
symmetry is broken.
In this case, the quasiparticle spectra are given by
\begin{eqnarray}
&&E({\bm k})
\nonumber\\
&=&\pm\sqrt{\left(\epsilon({\bm k})\pm |{\bm g}({\bm k})|\right)^2
+\left({\rm Re}\psi({\bm k})\pm |{\bm d}({\bm k})|\right)^2
+\left({\rm Im}\psi({\bm k})\right)^2
}.
\nonumber\\ 
\end{eqnarray}
If ${\rm Im}\psi({\bm k})$ is a nonzero constant, 
the line nodes in the mixed singlet-triplet states vanish completely.

\section{Conclusions and Discussions}
\label{sec:conclusion}

(1) First we would like to outline the results of this paper. We
examined topological stability of line nodes in superconducting states
with time-reversal invariance. 
A generic Hamiltonian of time-reversal invariant superconducting states
was presented. 
We found that only line nodes are topologically stable in single-band
descriptions of superconductivity.
It was shown that line nodes in high-$T_{\rm c}$ materials and mixed
singlet-triplet superconducting states are topologically stable, while
one in the polar state is not.
Using the time-reversal symmetry, we introduced a real structure
and defined ${\bm Z}_2$ topological numbers.
Stability of line nodes was explained by conservation of the
topological numbers.

(2) Besides the superconducting states examined in this paper, several
   superconductors such as ${\rm CeCoIn}_5$ \cite{TPNHBHRSTPCF}, ${\rm CeIrIn}_5$
   \cite{MJTPFPS} and ${\rm Sr_2RuO_4}$ \cite{STKMMI, ITYMSSFYSO,
   DMYM,DMM} are believed to host line nodes.
Among them, the gap function of ${\rm CeCoIn}_5$ is a $d$-wave paring,
   thus its line nodes are topologically stable in a similar manner as
   high-$T_{\rm c}$ materials. 
On the other hand, the line nodes of ${\rm Sr_2RuO_4}$ are not
   topologically stable since its superconducting state breaks the
   time-reversal symmetry \cite{LFKLMNUMMMNS}.

(3) An analysis based on the KR theory \cite{Horava} shows that point
nodes have a ${\bm Z}_2$ topological number in a nonrelativistic Fermi
system with the time-reversal symmetry. However, our analysis in
Sec.\ref{sec:hamiltonian} indicates that point nodes
have only the trivial topological number if the superconducting state
is described by single-band electrons. 

(4) If the superconductivity occurs in multi-bands of electrons,
a topologically stable point node is possible to exist. In other
words, if the time-reversal symmetry is not broken, the existence of a
topological stable point node implies a multi-band superconductivity.

(5) For mixed singlet-triplet superconducting states, the existence of
    the line nodes is not explained by the group theoretical
    method in Refs.\cite{VG,VG2,Blount, UR, UR2}. The topological stability
    studied here is essential for the existence of these line nodes.

(6) In addition to the time-reversal symmetry, if the party is
    conserved, the Hamiltonian becomes
\begin{eqnarray}
H({\bm k})=
\left(
\begin{array}{cc}
\epsilon({\bm k}) & i\psi({\bm k})\sigma_2\\
-i\psi({\bm k})\sigma_2 & -\epsilon({\bm k})
\end{array}
\right). 
\end{eqnarray}
This is a real symmetric matrix, thus we have a natural real structure.
Using this real structure, we can introduce a ${\bm Z}_2$ topological
number which is different from that given in
Sec.\ref{sec:topological_number} as follows \cite{Sato}.
Let us consider the eigenfunction of $H({\bm k})u({\bm k})=E({\bm
k})u({\bm k})$,
\begin{eqnarray}
u({\bm k})=
\left(
\begin{array}{c}
a({\bm k})\\
b({\bm k})\\
c({\bm k})\\
d({\bm k})
\end{array}
\right),
\end{eqnarray}
Here we can demand $a({\bm k})$, $b({\bm k})$, $c({\bm k})$ and $b({\bm
k})$ to be real since $H({\bm k})$ is a real matrix.
If we impose the normalization condition on $u({\bm k})$, we have $a({\bm
k})^2+b({\bm k})^2+c({\bm k})^2+d({\bm k})^2=1$, and by using the gauge
freedom of the eigenequation, $u({\bm k})$ can be identified with
$-u({\bm k})$. 
Therefore, $u({\bm k})$ is given by an element of $S^3/{\bm Z}_2$
\footnote{A more careful analysis shows that $u({\bm k})\in S^1/{\bm
Z}_2$, but in the following argument we do not need this.}.

Now consider a map determined by $u({\bm k})$ from an infinitesimal
$S^1$ around a line node into $S^3/{\bm Z}_2$. 
The map is classified by $\pi_1(S^3/{\bm Z}_2)$.
Using the homotopy theory \cite{Schwarz}, we obtain
\begin{eqnarray}
\pi_1(S^3/{\bm Z}_2)=\pi_0({\bm Z}_2)={\bm Z}_2. 
\end{eqnarray}
Therefore, we have a ${\bm Z}_2$ topological number corresponding to
$\pi_1(S^3/{\bm Z}_2)$.
The line node is topologically stable if the map
corresponds to the nontrivial element of ${\bm Z}_2$.
A straightforward calculation shows that the map around a line node in
high-$T_{\rm c}$ materials gives the nontrivial element. 
This is another explanation of topological stability of the line nodes
in high-$T_{\rm c}$ materials.

Although the construction of this topological number is easier than that
given in Sec.\ref{sec:topological_number}, it is available only when the
parity is conserved.

(7) We have noticed that when a line node splits in two as was shown in
    Sec.\ref{sec:ex_high_tc}, there remains a degenerate point between
    $E_{+}({\bm k})$ and $E_{+}({\bm k})$ near the line node. 
    The degenerate point is topologically stable since it is given by
    an intersection point of the three equations $\epsilon({\bm k}){\bm g}({\bm k})+\psi({\bm
    k}){\bm d}({\bm k})=0$ in three dimensional momentum space.
The degenerate point has a topological number similar to that of the
    chiral fermion in Ref.\cite{NN}. 
In order to describe the splitting of the line node in terms of the
    topological numbers given in this paper, we need to take into
    account the conservation law of this also.

\vspace{3ex}
Note added.- After completing this work, the author noticed a preprint
\cite{Volovik4} by G. E. Volovik in which the topological stability
of line nodes in high-$T_{\rm c}$ superconductors was discussed
very recently.
There is some overlap between his paper and
Sec.\ref{sec:topological_number_high_tc} in this paper.
His argument was restricted to the case where
the Hamiltonian is real, while the argument described here is not
restricted. 

\begin{acknowledgments}
I would like to express my gratitude to Mahito Kohmoto for stimulating
 discussions and encouragement. It is also a pleasure to thank Jun
 Goryo, Koich Izawa, Yuji Matsuda and Yong-Shi Wu for useful discussions.
\end{acknowledgments}

\newpage %Just because of unusual number of tables stacked at end
\bibliography{paper}% Produces the bibliography via BibTeX.

\end{document}